\documentclass[12pt,a4paper,titlepage]{article}

\usepackage{indentfirst} 
\usepackage{lastpage} 

\usepackage{amsmath} 
\usepackage{amssymb}
\usepackage{amsfonts}
\usepackage{amsthm}
\usepackage{bm} 
\usepackage{icomma} 

\usepackage{extsizes} 
\usepackage{geometry} 
\usepackage{setspace}
\usepackage{textcase}

\usepackage{cancel}

\usepackage[linkcolor=blue,colorlinks=true]{hyperref}

\newtheorem{lemma}{Lemma}
\newtheorem{theorem}{Theorem}

\theoremstyle{definition}
\newtheorem*{demo}{Proof}

\usepackage{epsfig}
\usepackage{graphicx}

\begin{document}
\begin{center}	
	
	\Large
	\textbf{Dynamics of moments of arbitrary order for stochastic Poisson squeezings}
		
	\large 
	\textbf{A.E. Teretenkov}\footnote{Department of Mathematical Methods for Quantum Technologies,
		Steklov Mathematical Institute of Russian Academy of Sciences, Moscow, Russia;\\
		Faculty of Physics, Lomonosov Moscow State University.\\ E-mail:taemsu@mail.ru}
	
\end{center}

\footnotesize
The explicit dynamics of the moments for the GKSL equation is obtained. In our case the GKSL equation corresponds to Poisson stochastic processes which lead to unitary jumps. We consider squeeze operators as the unitary jumps. \\


\normalsize

\section{Introduction}

In this work we consider the equation for the density matrix
\begin{equation}\label{eq:mainEq}
\frac{d}{dt} \rho_t = \mathcal{L}(\rho_t), \qquad \mathcal{L}(\rho) = \sum_{k=1}^K \lambda_k (U_k \rho U_k^{\dagger} - \rho), \qquad \lambda_k >0,
\end{equation}
where $ U_k $ are unitary operators with the generators which are quadratic in bosonic creation and annihilation operators (squeeze operators). Such generators naturally arise in the case of averaging with respect to classical Poisson processes. These processes have the intensities $ \lambda_k $ and lead to unitary jumps $ U_k $ \cite{Holevo96, Holevo98}. In a finite dimensional Hilbert space the dilation of form \eqref{eq:mainEq} by the Poisson process was discussed in \cite{Kummerer87}. Let us note that Poisson processes and the correspondent quantum Markov equation arise in physical applications \cite{accardi2002quantum, vacchini2009quantum, Basharov2014, TrubBash2018}. Unitary evolution with the quadratic generators mentioned above was discussed in \cite{Fried1953, Ber86, Manko79, Manko87, dodonov2003theory, Cheb11, Cheb12}. Let us also note that the generator $ \mathcal{L} $  has \textit{Gorini-Kossakowski-Sudarshan-Lindblad} (GKSL) form \cite{gorini1976completely, lindblad1976generators}
\begin{equation*}
\mathcal{L}(\rho) =  \sum_{k=1}^K \left(L_k \rho_t L_k^{\dagger} - \frac12 L_k^{\dagger} L_k\rho- \frac12 \rho L_k^{\dagger} L_k\right),
\end{equation*}
if one assumes $ L_k = \sqrt{\lambda_k} U_k $. 

We need an additional bit of notation to formulate our result. Notation here is similar to \cite{Ter16, Ter17a, Ter19}. We consider the Hilbert space $\otimes_{j=1}^n\ell_2$.  In such a space one could \cite[Paragraph 1.1.2]{scalli2003} define $n$ pairs of creation and annihilation operators satisfying \textit{canonical commutation relations} (CCR): $ [\hat{a}_i, \hat{a}_j^{\dagger}] = \delta_{ij}$, $ [\hat{a}_i, \hat{a}_j] = [\hat{a}_i^{\dagger}, \hat{a}_j^{\dagger}]= 0 $.  Let us define the $2n$-dimensional vector $\mathfrak{a} = (\hat{a}_1, \cdots, \hat{a}_n, \hat{a}_1^{\dagger}, \cdots, \hat{a}_n^{\dagger} )^T$.  Linear and quadratic forms in such operators we denote by $ f^T \mathfrak{a}  $ and $ \mathfrak{a}^T K \mathfrak{a} $, respectively.  Here, $  f \in \mathbb{C}^{2n} $ and $ K \in \mathbb{C}^{2n \times 2n} $.  Define the $2n \times 2n$-dimensional matrices as
\begin{equation*}
J = \biggl(
\begin{array}{cc}
0 & -I_n \\ 
I_n & 0
\end{array} 
\biggr), \qquad
E = \biggl(
\begin{array}{cc}
0 & I_n \\ 
I_n & 0
\end{array} 
\biggr),
\end{equation*}
where $ I_n $ is the identity matrix from $ \mathbb{C}^{n \times n} $. CCR in such a notation takes the form $ [f^T \mathfrak{a},\mathfrak{a}^T g]  = - f^T J g,  \forall g, f \in \mathbb{C}^{2n}, $ we also write it in the shorter form $ [\mathfrak{a},\mathfrak{a}^T]  = - J $. We also define the $\sim$-conju\-ga\-tion of vectors and matrices by the formulae
\begin{equation*}
\tilde{g} = E\overline{g}, \; g \in \mathbb{C}^{2n}, \qquad \tilde{K} = E \overline{K} E, \; K \in \mathbb{C}^{2n \times 2n},
\end{equation*}
where the overline is an (elementwise) complex conjugation. 

\begin{theorem}\label{th:main}
	Let the density matrix $ \rho_t $ satisfy Eq.~\eqref{eq:mainEq}, where the unitary operators $ U_k$, $ k=1, \ldots, K $, are defined by the formulae $ U_k = e^{- \frac{i}{2} \mathfrak{a}^T H_k \mathfrak{a} } $, $ H_k = H_k^T = \tilde{H}_k \in \mathbb{C}^{2n \times 2 n} $, and $  \langle \otimes_{m=1}^M \mathfrak{a} \rangle_0 < \infty$,  then the dynamics of the moments have the form
	\begin{equation}\label{eq:momDynam}
	\langle \otimes_{m=1}^M \mathfrak{a}   \rangle_t = e^{\sum_{k=1}^K\lambda_k (\otimes_{m=1}^M S_k  -  I_{(2n)^M}) t} \langle \otimes_{m=1}^M \mathfrak{a} \rangle_0, \qquad S_k = e^{i J H_k},
	\end{equation}
	where the average is defined by the formula $ \langle \otimes_{m=1}^M \mathfrak{a} \rangle_t \equiv \mathrm{tr} \; (\rho_t \otimes_{m=1}^M \mathfrak{a} ) $.
	In particular, for the first and second moments we have
	\begin{equation*}
	\langle\mathfrak{a}\rangle_t = e^{\sum_{k=1}^K \lambda_k(S_k  -  I_{2n}) t} \langle\mathfrak{a}\rangle_0, \qquad \langle\mathfrak{a} \otimes \mathfrak{a}\rangle_t = e^{\sum_{k=1}^K \lambda_k (S_k \otimes S_k -  I_{4 n^2}) t} \langle \mathfrak{a} \otimes \mathfrak{a} \rangle_0.
	\end{equation*}
\end{theorem}

Here $ I_{(2n)^m} $ is the identity matrix in $ \mathbb{C}^{2n} \otimes  \cdots \otimes \mathbb{C}^{2n}  = \mathbb{C}^{(2n)^m} $.  $  \langle \otimes_{m=1}^M \mathfrak{a} \rangle_0 < \infty$ means that the operators in the tensor $  \otimes_{m=1}^M \mathfrak{a} \rho_0 $ are nuclear.

\section{Moments dynamics} 

In this section we prove theorem \ref{th:main} splitting it into several lemmas.

\begin{lemma}
	\label{lem:conjGen}
	Let $ \rho $ be a nuclear operator and $ \hat{X} $ be an operator in $ \otimes_{j=1}^n\ell_2 $ which could be unbounded, but the operators $  U_k^{\dagger} \hat{X} U_k \rho  $, $ k=1, \ldots, K $ and $ \hat{X} \rho $ are nuclear, then
	\begin{equation*}
	\mathrm{tr} \,  \hat{X} \mathcal{L}(\rho) = 	\mathrm{tr} \, \mathcal{L}^*  (\hat{X})\rho,  
	\end{equation*}
	where
	\begin{equation}\label{eq:conjGen}
	\mathcal{L}^* (\hat{X}) = \sum_{k=1}^K \lambda_k (U_k^{\dagger} \hat{X} U_k - \hat{X}).
	\end{equation}
\end{lemma}

\begin{demo}
	As the trace of the nuclear operator is basis independent, then we have $ \mathrm{tr} \,(U_k^{\dagger} \hat{X} U_k \rho ) =  \mathrm{tr} \,(V U_k^{\dagger} \hat{X} U_k \rho  V^{\dagger})$ for an arbitrary unitary operator $ V $ in $ \otimes_{j=1}^n\ell_2 $. Assume $ V = U_k $, then $ \mathrm{tr} \,(U_k^{\dagger} \hat{X} U_k \rho ) =  \mathrm{tr} \,(\hat{X} U_k \rho U_k^{\dagger})$. By applying this reasoning to each summand \eqref{eq:mainEq} we obtain \eqref{eq:conjGen}. \qed
\end{demo}

Thus, the GKSL equation in the Heisenberg representation takes the form
\begin{equation}\label{eq:GKSLHeis}
\frac{d}{dt} \hat{X}_t = \sum_{k=1}^K \lambda_k (U_k^{\dagger} \hat{X} U_k - \hat{X})_t.
\end{equation}

Lemma 3.1 from \cite{Ter17a} in the case, when $ K = i H $, $ M =0 $, $ g=0 $, by the arbitrariness of $ f $ takes the following form.

\begin{lemma} 
	Let $ H = H^T \in \mathbb{C}^{2 n \times 2n} $, then
	\begin{equation*}
	e^{ \frac{i}{2} \mathfrak{a}^T H \mathfrak{a} }  \mathfrak{a} e^{- \frac{i}{2} \mathfrak{a}^T H \mathfrak{a} } = S \mathfrak{a} , \qquad	S = e^{i J H}.
	\end{equation*}
\end{lemma}

Let us note that in accordance with \cite{Ter16} in the case, when $ \tilde{H} = H$, the operator $ \frac12 \mathfrak{a}^T H \mathfrak{a} $ is self-adjoint. Hence, the operator $ e^{- \frac{i}{2} \mathfrak{a}^T H \mathfrak{a} }  $ is unitary. Thus, the operators $ U_k $ defined in theorem \ref{th:main} are unitary.

\begin{lemma}
	\label{lem:sympTransform}
	If $ \mathcal{L}^* $ is defined by formula \eqref{eq:conjGen} in the case, when $ U_k = e^{- \frac{i}{2} \mathfrak{a}^T H_k \mathfrak{a} } $, $ H_k = \tilde{H}_k \in \mathbb{C}^{2n \times 2 n} $, one has
	\begin{equation*}
	\mathcal{L}^*(\otimes_{m=1}^M \mathfrak{a}) =  \sum_{k=1}^K  \lambda_k (\otimes_{m=1}^M  S_k - I_{(2n)^M}) \otimes_{m=1}^M \mathfrak{a} , \qquad  S_k = e^{i J H_k}.
	\end{equation*}
\end{lemma}

\begin{demo}
	Taking into account lemma \eqref{eq:conjGen}
	\begin{equation*}
	\mathcal{L}^*(\otimes_{m=1}^M \mathfrak{a} ) 
	=  \sum_{k=1}^K \lambda_k (U_k^{\dagger} (\otimes_{m=1}^M \mathfrak{a} ) U_k - \otimes_{m=1}^M \mathfrak{a} ) = \sum_{k=1}^K \lambda_k (  \otimes_{m=1}^M (U_k^{\dagger} \mathfrak{a} U_k)  - \otimes_{m=1}^M \mathfrak{a}). 
	\end{equation*}
	By lemma \ref{lem:sympTransform} we have $ U_k^{\dagger} \mathfrak{a} U_k=  e^{\frac{i}{2} \mathfrak{a}^T H_k \mathfrak{a} } \mathfrak{a}  e^{- \frac{i}{2} \mathfrak{a}^T H_k \mathfrak{a} } = e^{i J H_k} \mathfrak{a} = S_k \mathfrak{a} $.
	Thus, we obtain
	\begin{equation*}
	\mathcal{L}^*(\otimes_{m=1}^M \mathfrak{a} ) 
	= \sum_{k=1}^K \lambda_k (  \otimes_{m=1}^M ( S_k \mathfrak{a} )  - \otimes_{m=1}^M \mathfrak{a}) =  \sum_{k=1}^K  \lambda_k (\otimes_{m=1}^M  S_k - I_{(2n)^M}) \otimes_{m=1}^M \mathfrak{a}.\qed
	\end{equation*}
\end{demo}

\noindent\textbf{Proof of theorem \ref{th:main}}.
	By lemmas \ref{lem:conjGen} and \ref{lem:sympTransform} we obtain the following GKSL eq\-ua\-tion in the Heisenberg representation \eqref{eq:GKSLHeis} in the case, when $ \hat{X}_t = (\otimes_{m=1}^M \mathfrak{a})_t $.
	\begin{equation*}
	\frac{d}{dt}  (\otimes_{m=1}^M \mathfrak{a})_t  = \sum_{k=1}^K \lambda_k (\otimes_{m=1}^M S_k - I_{(2n)^M})( \otimes_{m=1}^M \mathfrak{a})_t.
	\end{equation*}
	When we apply lemma \ref{lem:conjGen}, we take into account that the linear combinations of nuclear operators in the tensor $ \otimes_{m=1}^M \mathfrak{a} \rho_0 $ are also nuclear ones. The obtained equation is a linear ordinary differential equation with respect to the tensor $ ( \otimes_{m=1}^M \mathfrak{a})_t $. Hence, its solution could be represented in terms of matrix exponential of the $ (2n)^M\times(2n)^M $-matrix $ \sum_{k=1}^K \lambda_k (\otimes_{m=1}^M S_k - I_{(2n)^M}) t $, i.e.
	\begin{equation*}
	(\otimes_{m=1}^M \mathfrak{a})_t = e^{\sum_{k=1}^K \lambda_k (\otimes_{m=1}^M S_k - I_{(2n)^M})  t} (\otimes_{m=1}^M \mathfrak{a})_0.
	\end{equation*}
	By averaging over the initial density matrix $ \rho_0 $, we obtain \eqref{eq:momDynam}.\qed

\section{Conclusions}

In this work the explicit expressions for dynamics of the density matrix moments were obtained. This density matrix satisfies equation \eqref{eq:mainEq} with the squeeze operators. The possible directions of the further generalizations are analogous calculations in the fermionic case (it could be done by means of \cite{Ter19,Ter17}) and the consideration of arbitrary Gaussian channels \cite{Holevo15} instead of weighted sums of unitary Gaussian channels. In the latter case we also obtain linear differential equations for the moments, but they have more complicated form.

\end{document}